\documentclass[prd,amsmath,amssymb,superscriptaddress,showkeys,showpacs,twocolumn,10pt]{revtex4-1}

\pdfoutput=1

\usepackage{graphicx}
\usepackage{dcolumn}
\usepackage{bm}
\usepackage{amssymb}
\usepackage{latexsym}
\usepackage{booktabs}
\usepackage{amsmath}
\usepackage{multirow}
\usepackage{url}
\usepackage{subfigure}

\usepackage{float}
\usepackage[colorlinks=true, linkcolor=red, citecolor=blue]{hyperref}

\begin{document}
\title{Forecast for cosmological parameter estimation with gravitational-wave standard sirens from the LISA-Taiji network}

\author{Ling-Feng Wang}
\affiliation{Department of Physics, College of Sciences, Northeastern
University, Shenyang 110819, China}
\author{Shang-Jie Jin}
\affiliation{Department of Physics, College of Sciences, Northeastern
University, Shenyang 110819, China}
\author{Jing-Fei Zhang}
\affiliation{Department of Physics, College of Sciences, Northeastern
University, Shenyang 110819, China}

\author{Xin Zhang\footnote{Corresponding author}}
\email{zhangxin@mail.neu.edu.cn}
\affiliation{Department of Physics, College of Sciences, Northeastern
University, Shenyang 110819, China}
\affiliation{Key Laboratory of Data Analytics and Optimization
for Smart Industry (Northeastern University), Ministry of Education, Shenyang 110819, China}

\keywords{gravitational wave, space-based, network}

\begin{abstract}
LISA and Taiji are expected to form a space-based gravitational-wave (GW) detection network in the future. In this work, we make a forecast for the cosmological parameter estimation with the standard siren observation from the LISA-Taiji network. We simulate the standard siren data based on a scenario with configuration angle of $40^{\circ}$ between LISA and Taiji. Three models for the population of massive black hole binary (MBHB), i.e., pop III, Q3d, and Q3nod, are considered to predict the events of MBHB mergers.
We find that, based on the LISA-Taiji network, the number of electromagnetic (EM) counterparts detected is almost doubled compared with the case of single Taiji mission. Therefore, the LISA-Taiji network's standard siren observation could provide much tighter constraints on cosmological parameters. For example, solely using the standard sirens from the LISA-Taiji network, the constraint precision of $H_0$ could reach $1.3\%$. Moreover, combined with the CMB data, the GW-EM observation based on the LISA-Taiji network could also tightly constrain the equation of state of dark energy, e.g., the constraint precision of $w$ reaches about $4\%$, which is comparable with the result of CMB+BAO+SN. It is concluded that the GW standard sirens from the LISA-Taiji network will become a useful cosmological probe in understanding the nature of dark energy in the future.
%
%
\end{abstract}

\pacs{95.36.+x, 98.80.Es, 98.80.-k, 04.80.Nn, 95.55.Ym}
\keywords{LISA-Taiji network, gravitational-wave standard siren, cosmological parameter estimation, massive black bole binary, electromagnetic counterpart}

\maketitle

\section{Introduction}\label{sec:intro}

The precise measurement of the cosmic microwave background (CMB) anisotropies initiated the era of precision cosmology \cite{Spergel:2003cb,Bennett:2003bz}.
Constraining the standard cosmological model, i.e., the $\Lambda$CDM model, with the high-precision CMB observations enables cosmologists to have a comprehensive understanding of the evolution history of the universe. However, the accurate measurements also led to some puzzling issues. For example, there is a 4.4$\sigma$ tension between the $H_{0}$ values inferred from the CMB observation \cite{Aghanim:2018eyx} and the distance ladder measurement \cite{Riess:2019cxk}.
Essentially, the $H_0$ tension reflects an inconsistency of measurements between the early universe and the late universe \cite{Verde:2019ivm,Riess:2020sih}. In addition to facing the challenge of the Hubble tension, the $\Lambda$CDM model also has some theoretical problems, such as the ``fine-tuning" and ``cosmic coincidence" problems \cite{Weinberg:1988cp,Sahni:1999gb,Bean:2005ru}, which implies that the $\Lambda$CDM model needs to be further adjusted.
Therefore, the current development of cosmology can be divided into two main aspects: (i) further extending the standard $\Lambda$CDM model \cite{Guo:2018ans,Zhao:2018fjj,DiValentino:2021izs,Li:2018ydj,Vagnozzi:2019ezj,Guo:2018uic,Li:2020gtk,Feng:2019jqa,Zhang:2020mox,Feng:2019mym,Lin:2020jcb,Guo:2019dui,Hryczuk:2020jhi,Zhang:2019ipd,Gao:2021xnk}, and (ii) developing more low-redshift observation projects aimed at precisely measuring the late universe \cite{Zhang:2019ylr,Xu:2020uws,Li:2020tds,Zhang:2021yof,Wang:2021kxc}. For the second aspect, the gravitational-wave (GW) standard siren method \cite{Schutz:1986gp,Holz:2005df} is one of the most promising options and has been widely discussed \cite{Holz:2005df,Cai:2017buj,DiValentino:2017clw,Wei:2018cov,Zhao:2018gwk,DiValentino:2018jbh,Du:2018tia,Mifsud:2019fut,Wei:2019fwp,Gray:2019ksv,Howlett:2019mdh,Chassande-Mottin:2019nnz,Doctor:2019odr,Fu:2019oll,Mukherjee:2019qmm,Abbott:2019yzh,Palmese:2020aof,Chen:2020dyt,Wang:2020xwn,Chen:2020zoq,Zhao:2010sz,Cai:2016sby,Cai:2017aea,Yang:2017bkv,Cai:2017plb,Wang:2018lun,Zhang:2018byx,Li:2019ajo,Zhang:2019loq,Zhang:2019ple,Yan:2019sbx,Yang:2019vni,Jin:2020hmc,Qi:2021iic}.

The amplitude of GW generated by the merger of compact binary encodes luminosity distance and chirp mass of the source.
The absolute luminosity distance could be obtained if the amplitude is measured precisely, and the chirp mass could be inferred from the variation of GW frequency.
The relation between luminosity distance $d_{\rm L}$ and redshift $z$ can be established, once the electromagnetic (EM) counterpart of a GW source is detected by optical observatories. Since the $d_{\rm L}$--$z$ relation is determined by the expansion history of the universe, cosmological models could be constrained by this relation. This method is usually referred to as the ``standard siren" method \cite{Schutz:1986gp,Holz:2005df}. GW170817 \cite{TheLIGOScientific:2017qsa,GBM:2017lvd,Monitor:2017mdv}, the first detected binary neutron star (BNS) merger event with an EM counterpart (GRB 170817A), has provided an independent measurement of the Hubble constant, giving the result of $H_0=70^{+12}_{-8}~{\rm km}~{\rm s}^{-1}~{\rm Mpc}^{-1}$ \cite{Abbott:2017xzu}. Recently, the event ZTF19abanrhr \cite{Graham:2020gwr} reported by the Zwicky Transient Facility is regarded as a candidate of the first plausible optical EM counterpart to the binary black hole (BBH) merger event GW190521 \cite{Abbott:2020tfl}. Chen \emph{et al.} \cite{Chen:2020gek} and Mukherjee \emph{et al.} \cite{Mukherjee:2020kki} have given the constraints on the $\Lambda$CDM and $w$CDM models, assuming ZTF19abanrhr is the actual EM counterpart to GW190521. Even if BBH merger events are expected to have no EM counterparts, these ``dark sirens" could also be used in cosmological fits using the statistical method discussed in Refs.~\cite{DelPozzo:2012zz,Chen:2017rfc,Fishbach:2018gjp,Feeney:2018mkj,Ding:2018zrk,Soares-Santos:2019irc,Lagos:2019kds,Yu:2020vyy}.

The potential of standard siren method in constraining cosmological parameters has been forecasted in Refs.~\cite{Zhao:2010sz,Cai:2016sby,Cai:2017aea,Yang:2017bkv,Cai:2017plb,Wang:2018lun,Zhang:2018byx,Li:2019ajo,Zhang:2019loq,Zhang:2019ple,Yan:2019sbx,Yang:2019vni,Jin:2020hmc,Qi:2021iic}, based on future ground-based GW detectors, e.g., Einstein Telescope \cite{Punturo:2010zz,ET-web} and Cosmic Explorer \cite{Evans:2016mbw,CE-web}. Several mechanisms for producing fast radio bursts (FRBs) by the mergers of binaries, such as charged black holes or neutron stars, are proposed in Refs.~\cite{Totani:2013lia,Mingarelli:2015bpo,Wang:2016dgs,Liu:2016olx,Zhang:2016rli,Yamasaki:2017hdr}, and subsequently, GW/FRB association systems as a complementary cosmological probe are discussed in Refs.~\cite{Wei:2018cgd,Cai:2019cfw} (see also Ref.~\cite{Zhao:2020ole}).
The GW sources detected by future ground-based GW detectors are BNSs or stellar-mass BBHs, which are mainly distributed at $z<3$. {In addition, the GWs produced by the massive black hole binaries (MBHBs) with EM counterparts are also expected to serve as standard sirens \cite{Holz:2005df}. The MBHB standard sirens may be detected at the redshift up to $z\simeq10$ in the future \cite{Caldwell:2019vru}, providing a promising method of measuring the expansion history of the universe back to a much earlier time.}
Space-based GW detection missions have been proposed and implemented to detect the GWs produced by MBHBs.

The Laser Interferometer Space Antenna (LISA) \cite{LISA-web,Armano:2016bkm,Audley:2017drz,Armano:2018kix,Abich:2019cci,Speri:2020hwc} is a European space-based GW observatory, with three identical drag-free spacecraft that form an equilateral triangle with arm length of $2.5\times 10^6$ km. Taiji \cite{Wu:2018clg,Guo:2018npi,Hu:2017mde} and TianQin \cite{Luo:2020bls,Wang:2019tto,Liu:2020eko,Milyukov:2020kyg,Mei:2020lrl,Fan:2020zhy} are two space-based GW observatories proposed by Chinese researchers. Taiji is a LISA-like space-based GW observatory proposed by the Chinese Academy of Sciences, also with a triangle of three satellites but with arm length of $3\times 10^6$ km. Some forecasts for the capability of LISA and Taiji in cosmological parameter estimation have been discussed in Refs.~\cite{Tamanini:2016zlh,Belgacem:2019pkk,Zhao:2019gyk}.
Multiple GW observatories can constitute a network to improve the measurement precision of source parameters \cite{Abbott:2017oio,Fan:2018vgw}, by measuring phase differences and amplitude ratios of GWs in different detectors. LISA's orbit is proposed to be at the ecliptic plane behind the Earth with a $20^{\circ}$ trailing angle, and Taiji is planed to be localized in front of the Earth with a $20^{\circ}$ leading angle, so that LISA and Taiji could form a space-based network aimed at detecting GW signals within an mHz range (i.e. $10^{-4}$ Hz to $10^{-1}$ Hz) \cite{Ruan:2020smc,Wang:2020vkg,Hu:2020rub,Wang:2020dkc,Omiya:2020fvw,Orlando:2020oko,Wang:2021mou}.

Recently, the LISA-Taiji network's capability of localizing GW sources was detailedly discussed in Refs.~\cite{Ruan:2020smc,Wang:2020vkg}. Wang \emph{et al.} \cite{Wang:2020dkc} showed that within 5-year operation time, the LISA-Taiji network is able to constrain the Hubble constant within $1\%$ accuracy via dark sirens. Omiya and Seto \cite{Omiya:2020fvw} explored the detectability of vector and scalar polarization modes in a stochastic gravitational wave background (SGWB) around 1 mHz with the LISA-Taiji network. Orlando \emph{et al.} \cite{Orlando:2020oko} proposed that the chirality of an isotropic SGWB can be detected by cross-correlating the data streams of LISA and Taiji. Wang and Han \cite{Wang:2021mou} showed that the LISA-Taiji network could improve the observations on the anomalous polarization predicted by the theories beyond general relativity.

In this work, we focus on the LISA-Taiji network's capability of improving the constraint accuracies of cosmological parameters. We first use the Fisher information matrix method to estimate the uncertainty of luminosity distance $d_{\rm L}$, and then simulate 5-year standard siren data based on the LISA-Taiji network. Then, we constrain three typical dark-energy cosmological models, i.e., the $\Lambda$CDM, $w$CDM, and CPL models, using the standard siren (joint GW-EM detection) mock data. We mainly analyze the improvement on the constraint accuracy of equation of state (EoS) of dark energy, and show that the LISA-Taiji network will play an important role in exploring the properties of dark energy in the future.

The rest of this paper is organized as follows. {In Sec.~\ref{sec:GW}, we describe the GW waveform and the detector response. In Sec.~\ref{sec:Fisher}, we describe the Fisher matrix analysis for GW parameter estimation. In Sec.~\ref{sec:EM}, we discuss the identifications of EM counterparts. In Sec.~\ref{sec:catalog}, we introduce the methods of simulating the standard siren catalog. In Sec.~\ref{sec:Result}, we display the constraint results and make some discussions. The conclusion is given in Sec.~\ref{sec:con}.} Unless otherwise specified, we adopt the system of units in which $c=G=1$ throughout this paper.

\section{Simulation of standard siren observation}\label{sec:simulation}

\subsection{GW waveform and detector response}\label{sec:GW}

The GW signal from the inspiral of a non-spinning MBHB can be modeled by the restricted post-Newtonian (PN) waveform. The GW strain $h(t)$ can be described by two independent polarizations ${h_{+,\times}}(t)$ in the transverse-traceless gauge,
\begin{align}
 h(t) = {F_ + }(t;\theta ,\phi ,\psi ){h_ + }(t) + {F_ \times }(t;\theta ,\phi ,\psi ){h_ \times }(t)  \,,
\end{align}
where $F_{+,\times}$ are antenna pattern functions, $(\theta,\phi )$ denote the source's polar angle and azimuthal angle in the ecliptic frame, and $\psi $ is the polarization angle of GW. We can separate the antenna pattern function into a polarization angle part and a $D_{+,\times}$ part that describes the dependence of time,
\begin{align}
 F_{+}(t) =& \frac{1}{2}\Big({\rm cos}(2\psi)D_{+}(t)-{\rm sin}(2\psi)D_{\times}(t)\Big),  \label{Fplus}\\
 F_{\times}(t) =& \frac{1}{2}\Big({\rm sin}(2\psi)D_{+}(t)+{\rm cos}(2\psi)D_{\times}(t)\Big)  \label{Fcros}.
\end{align}

For the inspiral process, the specific forms of $D_{+,\times}$ with the low-frequency approximation are given in Ref.~\cite{Ruan:2020smc},
\begin{align}
D_{+}(t) =& \frac{\sqrt{3}}{64}\bigg[-36{\rm sin}^2\theta\,{\rm sin}\big(2\alpha(t)-2\beta\big)+\big(3+{\rm cos(2\theta)}\big)\nonumber \\
&\times\bigg({\rm cos}(2\phi)\Big(9\sin(2\beta)-{\rm sin}\big(4\alpha(t)-2\beta\big)\Big)\nonumber \\
&+{\rm sin}(2\phi)\Big({\rm cos}\big(4\alpha(t)-2\beta\big)-9\cos(2\beta)\Big)\bigg)\nonumber \\
&-4\sqrt{3}{\rm sin}(2\theta)\Big({\rm sin}\big(3\alpha(t)-2\beta-\phi\big)-3{\rm sin}\big(\alpha(t)\nonumber \\
&-2\beta+\phi\big)\Big)\bigg]  \,,\label{Dplus}  \\
D_{\times}(t) =& \frac{1}{16}\bigg[\sqrt{3}{\rm cos}\theta\Big(9{\rm cos}(2\phi-2\beta)-{\rm cos}\big(4\alpha(t)-2\beta\nonumber \\
&-2\phi\big)\Big)-6{\rm sin}\theta\Big({\rm cos}\big(3\alpha(t)-2\beta-\phi\big)\nonumber \\
&+3{\rm cos}\big(\alpha(t)-2\beta+\phi\big)\Big)\bigg]  \,,\label{Dcros}
\end{align}
where $\alpha=2\pi f_m t+\kappa$ is the orbital phase of the guiding center, and $\beta$ is the initial orientation of the constellation. We simply set $\beta=0$ in our simulation. The triangular GW detectors with three arms, such as LISA and Taiji, can be equivalent to two independent $90^{\circ}$-interferometers (i.e., ``L-shaped" interferometers). The second interferometer is equivalent to the first one rotated by $\pi/4$ radians. The response functions of the two interferometers are ${F_{+,\times}}(t;\theta ,\phi ,\psi )$ and ${F_{+,\times}}(t;\theta ,\phi-\pi/4,\psi )$ \cite{Cutler:1997ta}. Here $\kappa$ is the initial ecliptic longitude of the guiding center and $f_m=1/{\rm yr}$. We assume that $\kappa=0$ for LISA and $\kappa=40^{\circ}$ for Taiji, so that the separation angle between LISA and Taiji is $40^{\circ}$.


For the sake of describing GW signals in the Fourier space, the observation time $t$ in Eqs.~(\ref{Dplus}) and (\ref{Dcros}) is replaced by \cite{Krolak:1995md,Buonanno:2009zt}
\begin{equation}
   t(f) = t_{\rm c} - \frac{5}{256} M_{\rm c} ^{-5/3}(\pi f)^{-8/3},
\end{equation}
where $t_{\rm c}$ is the coalescence time of MBHB.
The Fourier transformation of the strain can be obtained, i.e.,
\begin{eqnarray}
\label{eq:strainf}
\tilde{h}(f)=-\left(\frac{5\pi}{24}\right)^{1/2}M_{\rm c}^{5/6}\left[\frac{(\pi f)^{-7/6}}{D_{{\rm eff}}}\right] e^{-i\Psi}.
\end{eqnarray}
The effective luminosity distance, $D_{\rm eff}$, is defined as
\begin{equation}
D_{{\rm eff}}=d_{\rm L} \left[F^{2}_{+}\left(\frac{1+{\rm cos}^2 \iota}{2}\right)^2+F^{2}_{\times} {\rm cos}^2 \iota
\right]^{-1/2},
\end{equation}
where $d_{\rm L}$ is the luminosity distance to a GW source and $\iota$ is the inclination angle between the orbital angular momentum and the line of sight. $\Psi$ can be written to the second PN order as
\begin{align}
\Psi(f;M_c,\eta) =& 2\pi ft_0-2\phi_0-\frac{\pi}{4}+\frac{3}{128\eta}\bigg[\nu^{-5}+\Big(\frac{3715}{756} \nonumber \\
&+\frac{55}{9}\eta\Big)\nu^{-3}-16\pi\nu^{-2}+\Big(\frac{15293365}{508032}\nonumber \\
&+\frac{27145}{504}\eta+\frac{3085}{72}{\eta}^2\Big)\nu^{-1}\bigg], \\
\nu =& \Big(\frac{G\pi M}{c^3}f\Big)^{1/3},
\end{align}
where $t_0=t_c+\tau(t)$ is the coalescence time at a detector. According to the forward modeling of LISA described in Ref.~\cite{Rubbo:2003ap}, to linear order in eccentricity, the time delay $\tau(t)$ and the phase $\phi_0$ take the forms
\begin{align}
\tau(t)=&-\frac{R}{c}{\rm sin}\theta\,{\rm cos}(\alpha -\phi)-\frac{1}{2}e\frac{R}{c}{\rm sin}\theta\big[\cos(2\alpha-\phi-\beta)\nonumber\\
&-3\cos(\phi-\beta)\big],
\end{align}
\begin{align}
 2\phi_0=2\phi_c-{\rm arctan}\left(\frac{F_{\times}(\theta,\phi,\iota,\psi;t)}{F_{+}(\theta,\phi,\iota,\psi;t)}\frac{2{\rm cos}\iota}{1+{\rm cos}^2 \iota}\right),
\end{align}
where $R = 1\ {\rm AU}$, and $e$ is the eccentricity of detector's orbit. In the observer's reference frame, $M_{\rm c} = (1 + z)\eta^{3/5}M$ is the redshifted chirp mass. $M = {M_1} + {M_2}$ is the total mass of MBHB with ${M_1}>{M_2}$, and ${\eta  = M_1M_2 /M^2 }$ is the symmetric mass ratio.

\subsection{Fisher matrix analysis for GW parameter estimation}\label{sec:Fisher}

For a network including $N$ independent detectors, the Fisher information matrix can be written as
\begin{equation}
\label{eq:fisherm}
\boldsymbol{F}_{ij}=\left(\frac{\partial \boldsymbol{h}(f)}{\partial \theta_i}\bigg | \frac{\partial \boldsymbol{h}(f)}{\partial \theta_j}\right),
\end{equation}
with $\boldsymbol{h}$ being given by
\begin{equation}
\boldsymbol{h}(f)=\left[\frac{\tilde{h}_1 (f)}{\sqrt{S_{\rm n} (f)}},\frac{\tilde{h}_2 (f)}{\sqrt{S_{\rm n} (f)}},\cdots,\frac{\tilde{h}_N (f)}{\sqrt{S_{\rm n} (f)}}\right]^{\rm T},
\end{equation}
where $\theta_i$ denotes nine parameters ($d_L$, $M_c$, $\eta$, $\theta$, $\phi$, $\iota$, $t_c$, $\phi_c$, $\psi$) for a GW event.
Here, $S_{\rm n} (f)$ is the noise power spectral density. The specific forms of $S_{\rm n} (f)$ for Taiji and LISA are obtained from Refs.~\cite{Guo:2018npi,Klein:2015hvg}.
The bracket in Eq.~\eqref{eq:fisherm} for two functions $a(t)$ and $b(t)$ is defined as
\begin{equation}
\label{eq:filter}
(a|b)=4\int_{f_{\rm low}}^{f_{\rm up}}\frac{\tilde{a}(f)\tilde{b}^{*}(f)
+\tilde{a}^{*}(f)\tilde{b}(f)}{2}{\rm d}f,
\end{equation}
where ``$\sim$" above a function denotes the Fourier transform of the function.
The upper limit of the integral is set to the innermost stable circular orbit (ISCO) frequency $f_{\rm ISCO}=c^3/(6\sqrt{6}\pi G M)$ ~\cite{Feng:2019wgq}, and the lower frequency cutoff is set to $f_{\rm low}=10^{-4}\,\rm Hz$.
The signal-to-noise ratio (SNR) of a GW event is given by
\begin{equation}
\rho^2=(\boldsymbol{h}|\boldsymbol{h}),
\end{equation}
and we consider the SNR threshold of 8 in our simulation.

The Fisher matrix of the LISA-Taiji network is the sum of the Fisher matrix of LISA and the one of Taiji, which can be expressed as
\begin{equation}
\label{eq:filter}
F_{{\rm network}}=F_{{\rm LISA}}+F_{{\rm Taiji}}.
\end{equation}
Then, the errors of GW parameters can be estimated by the Fisher information matrix,
\begin{equation}
\Delta \theta_i =\sqrt{(F^{-1})_{ii}}.
\end{equation}

In our analysis, we take into account nine parameters ($d_L$, $M_c$, $\eta$, $\theta$, $\phi$, $\iota$, $t_c$, $\phi_c$, $\psi$) in the Fisher matrix. The error of luminosity distance, $\Delta d_{\rm L}$, and the angular resolution, $\Delta\Omega$, could be calculated by the Fisher matrix. Here, $\Delta\Omega$ is given by $\Delta\Omega=2\pi|\textrm{sin}\theta|\sqrt{\langle\Delta\theta^{2}\rangle\langle\Delta\phi^{2}\rangle-\langle\Delta\theta\Delta\phi\rangle^{2} }$, with $\langle\Delta\theta^{2}\rangle$, $\langle\Delta\phi^{2}\rangle$, and $\langle\Delta\theta\Delta\phi\rangle$ being given by the inverse of the 9-parameter Fisher matrix \cite{Zhao:2017cbb}. Improving the angular resolutions of GW sources is helpful to identify EM counterparts, so $\Delta\Omega$ is important for determining the number of GW-EM events. In addition, $\Delta d_{\rm L}$ could directly affect the constraint accuracies of cosmological parameters. Therefore, before making a cosmological analysis, it is necessary to study the reductions of $\Delta\Omega$ and $\Delta d_{\rm L}$ made by the LISA-Taiji network.

\begin{figure}[h]
\centering
\includegraphics[width=1.05\linewidth,angle=0]{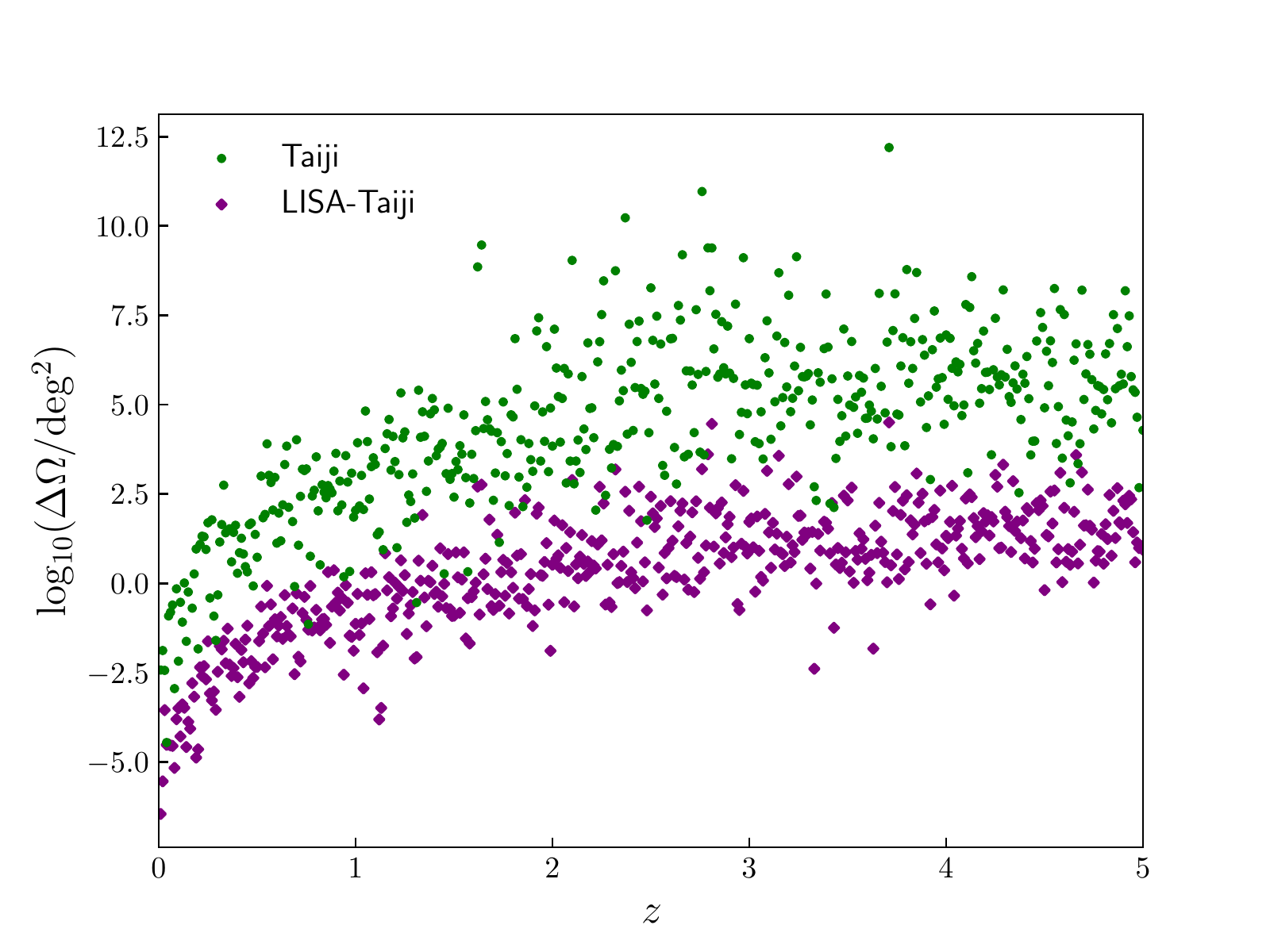}
\includegraphics[width=1.05\linewidth,angle=0]{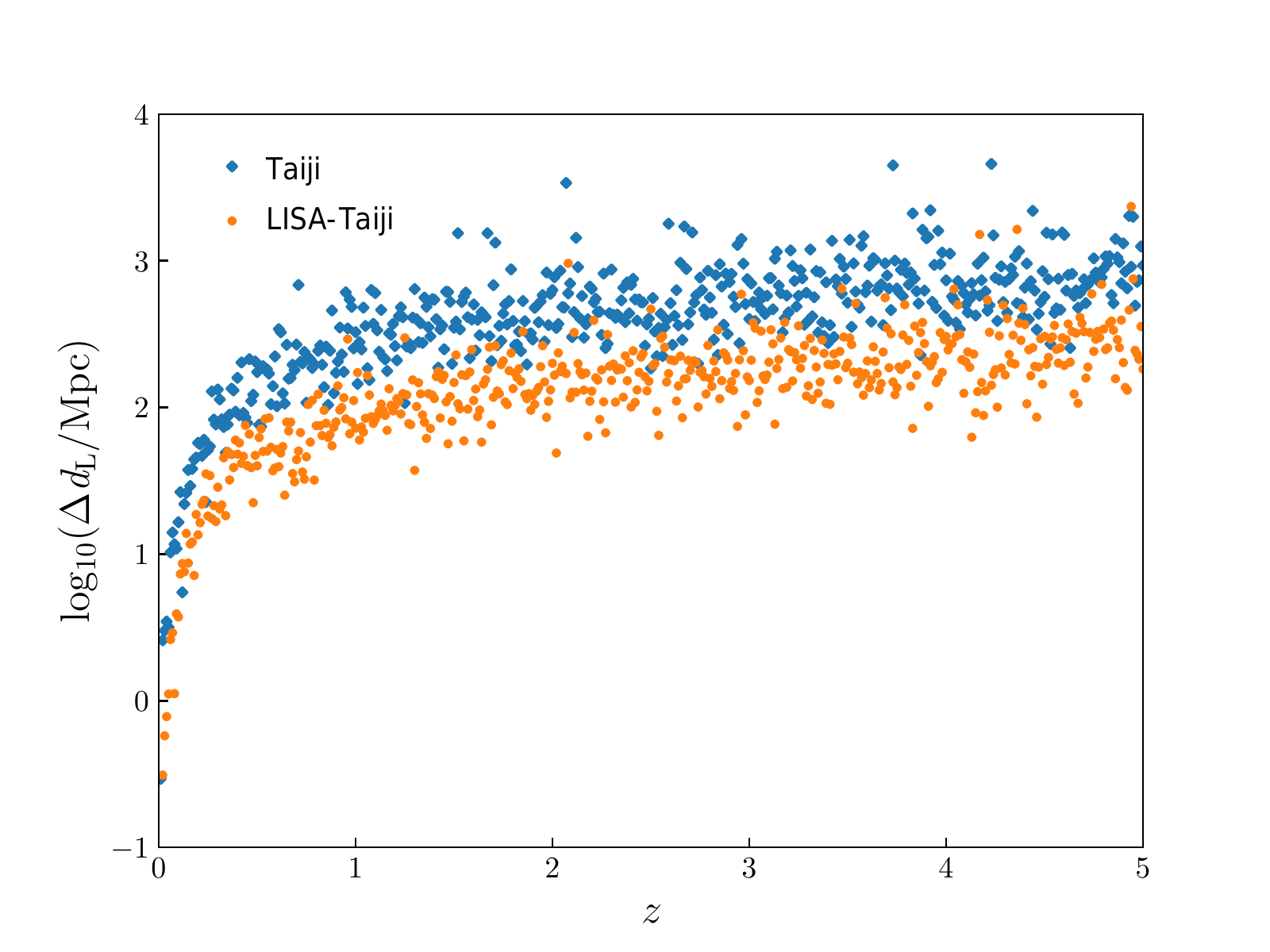}
\caption{The uncertainties of angular resolutions and luminosity distances as functions of redshift. Here we choose $\kappa=0$ for LISA, and $\kappa=40^{\circ}$ for Taiji.} \label{zdl}
\end{figure}

For the sake of clearly showing the reductions of $\Delta\Omega$ and $\Delta d_{\rm L}$, we plot $\Delta\Omega$ and $\Delta d_{\rm L}$ as functions of redshift in Figure \ref{zdl}. In this figure, we simulate 500 GW events to show statistical distributions of $\Delta\Omega$ and $\Delta d_{\rm L}$. We choose $\kappa=0$ for LISA and $\kappa=40^{\circ}$ for Taiji. The mass of MBHs, the sky location ($\theta$, $\phi$), the binary inclination $\iota$, the polarization angle $\psi$, and the coalescence phase $\phi_c$ are randomly chosen in the ranges of $[10^4, 10^7]M_{\odot}$, $[0,\pi]$, $[0,2\pi]$, $[0,\pi]$, $[0,2\pi]$, and $[0,2\pi]$, respectively. {We can clearly see that the LISA-Taiji network could reduce $\Delta\Omega$ by several orders of magnitude compared with the single Taiji mission, which implies that the LISA-Taiji network could greatly improve the capability of locating GW sources, and thus could increase the detection number of GW-EM events. For the uncertainty of luminosity distance, $\Delta d_{\rm L}$, it is not reduced as much as $\Delta\Omega$, but is still reduced by a factor of a few.} We have a more specific analysis on these aspects in Sec.~\ref{sec:catalog}.

Actually, $\Delta d_{\rm L}$ estimated by the Fisher matrix is the instrumental error, $\sigma_{d_{\rm L}}^{\rm inst}$. The total measurement error of luminosity distance $\sigma_{d_{\rm L}}$ also consists of the lensing error, the peculiar velocity error, and the redshift measurement error, which can be expressed as \cite{Wang:2019tto}
\begin{align}\label{eq:DECIGOlensingError}
(\sigma_{d_{\rm L}})^{2}=(\sigma_{d_{\rm L}}^{\rm inst})^{2}+(\sigma_{d_{\rm L}}^{\rm lens})^{2}+(\sigma_{d_{\rm L}}^{\rm pv})^{2}+(\sigma_{d_{\rm L}}^{\rm reds})^{2}.
\end{align}
The main systematic error caused by weak lensing is adopted from the fitting formula \cite{Tamanini:2016zlh},
\begin{align}\label{eq:DECIGOlensingError}
\sigma_{d_{\rm L}}^{\rm lens}(z)=d_{\rm L}(z)\times 0.066\bigg[\frac{1-(1+z)^{-0.25}}{0.25}\bigg]^{1.8},
\end{align}
and the error caused by the peculiar velocity of a source should also be included \cite{Kocsis:2005vv},
\begin{align}
\sigma_{d_{\rm L}}^{\rm pv}(z)=d_{\rm L}(z)\times\bigg[1+\frac{c(1+z)^2}{H(z)d_{\rm L}(z)}\bigg]\frac{\sqrt{\langle v^{2}\rangle}}{c},
\end{align}
where the peculiar velocity $\sqrt{\langle v^{2}\rangle}$ of the source with respect to the Hubble flow is roughly set to $500\,\mathrm{km\,s^{-1}}$.

The error from the redshift measurement of the EM counterpart could be ignored if the redshift is measured spectroscopically. But when using photometric redshift for a distant source, this factor should be taken into account. We estimate the error on the redshift measurement as
\begin{align}
\sigma_{d_{\rm L}}^{\rm reds}=\frac{\partial d_{\rm L}}{\partial z} (\Delta z)_n,
\end{align}
with $(\Delta z)_n\simeq 0.03(1+z_n)$ \cite{Ilbert:2013bf}.

\subsection{{Identifications of electromagnetic counterparts}}\label{sec:EM}
{In addition to the luminosity distance obtained from the GW waveform, the realization of standard siren also requires the redshift information.
Moreover, the number of GW-EM events has a great impact on the estimations of cosmological parameters, so we give an analysis on the EM counterpart below.}

{In the process of the merger of MBHB with external magnetic fields, it is assumed that EM radiations could be emitted in both the radio and optical bands \cite{Palenzuela:2010nf,OShaughnessy:2011nwl,Moesta:2011bn,Kaplan:2011mz,Shi:2011us,Blandford:1977ds,Meier:2000wk,Dotti:2011um}. The radiation in the radio frequency band can be detected using SKA \cite{SKA-web}, to identify the host galaxy of the GW source. Subsequently, the radiation in the optical band can be measured spectroscopically or photometrically through optical/IR projects, such as the Vera C. Rubin Observatory \cite{Ivezic:2008fe} and the European Extremely Large Telescope (E-ELT) \cite{ELT-web}, to obtain the redshift information.}

In practice, before the EM projects make observations, it is necessary to accurately locate the GW source.  Because the fields of view of the EM projects, such as SKA and ELT, are about 10 ${\rm deg}^2$, in our simulation, we choose those GW events with $\Delta\Omega<10~{\rm deg}^2$. From our analyses in Sec.~\ref{sec:Fisher}, one can see that the most obvious advantage of a GW detection network over a single detector is that it can greatly improve the capability of locating GW sources, which is also discussed in Ref.~\cite{Ruan:2020smc}.
It should be mentioned that we use a conservative scenario to estimate the location parameter by using only the inspiral phase \cite{Tamanini:2016zlh}. A more optimistic scenario is to include the merger and ringdown phases, which could lead to more GW-EM events. We leave this issue for future research.

After locating the GW source within $10~{\rm deg}^2$ by the GW detectors, we need to further uniquely identify the host galaxy by the EM counterpart. Of course, if the host galaxy cannot be uniquely identified, statistical methods can also be applied in estimating cosmological parameters. We leave the relevant discussion in the future work. To simulate the EM counterpart, it is necessary to understand the formation mechanism of MBHB and its external environment. Regarding these aspects, different theoretical models have been proposed \cite{Klein:2015hvg}. In this paper, we use an analysis method similar to that used in Refs.~\cite{Tamanini:2016zlh,Yang:2021qge}. Next, we briefly introduce this method. More details can be found in Ref.~\cite{Tamanini:2016zlh}.

We first simulate the EM radiation in the radio band, which will be used to uniquely identify the host galaxy. The total luminosity $L_{\rm radio}$ in the radio band consists of two parts.
One part is the dual jet $L_{\rm flare}$ emitted when a binary is close to merging, which is caused by the twisting of external magnetic field lines by the rapidly inspiralling MBHB \cite{Palenzuela:2010nf,OShaughnessy:2011nwl,Moesta:2011bn,Kaplan:2011mz,Shi:2011us}, being given by
\begin{align}
L_{\rm flare}=\epsilon_{\rm edd}\epsilon_{\rm radio}(v/v_{\rm max})^2q^2L_{\rm edd}.
\end{align}
The factor $(v/v_{\rm max})^2$ describes the luminosity evolution as the MBHB inspirals ($v_{\rm max}=c/\sqrt{3}$ is the circular speed at the innermost stable circular orbit for a binary of BHs, and $v$ is the binary's coordinate relative circular velocity \cite{OShaughnessy:2011nwl}). $\epsilon_{\rm radio}$ is the fraction of EM radiations emitted in the radio band (i.e., a radio-to-bolometric luminosity correction), and is set to a fiducial value of 0.1. $\epsilon_{\rm edd}$ is the Eddington ratio, which is calculated according to the formulas shown in Appendix A of Ref.~\cite{Tamanini:2016zlh}. $q = M_2/M_1 \leq 1$ is the binary's mass ratio.
The other part of $L_{\rm radio}$ is the standard radio jet due to the Blandford-Znajeck effect \cite{Blandford:1977ds,Meier:2000wk}, with luminosity dependent on the mass accretion rate. Following Ref.~\cite{Tamanini:2016zlh}, we use the jet luminosity
\begin{widetext}
\begin{align}
L_{\rm jet}=\begin{cases}10^{42.7}{\rm erg}~{\rm s}^{-1}(\frac{\alpha}{0.01})^{-0.1}m_9^{0.9}(\frac{\dot{m}}{0.1})^{6/5}(1+1.1a_1+0.29a_1^2),~{\rm if}~10^{-2}\leq\epsilon_{\rm edd}\leq0.3, \\
10^{45.7}{\rm erg}~{\rm s}^{-1}(\frac{\alpha}{0.3})^{-0.1}m_9(\frac{\dot{m}}{0.1})g^2(0.55f^2+1.5fa_1+a_1^2),~{\rm otherwise.}
\end{cases}
\end{align}
\end{widetext}
We assume the Shakura-Sunyev viscosity parameter $\alpha = 0.1$; $m_9$ is defined as $m_9 = M_1/(10^9M_\odot)$; $\dot{m}$ is the central accretion rate, which is calculated from Appendix A of Ref.~\cite{Tamanini:2016zlh}; $a_1$ is the spin parameter of the BH with the mass of $M_1$; $f$ and $g$ are dimensionless quantities regulating the angular velocity and the azimuthal magnetic field, respectively, and are set to $f = 1$ and $g = 2.3$ \cite{Meier:2000wk}. The total luminosity in the radio band is given by $L_{\rm radio}=L_{\rm flare}+L_{\rm jet}$.
For the GW event satisfying $L_{\rm radio}\geq4\pi d_{\rm L}^2 F^{\rm SKA}_{\rm min}$ \cite{OShaughnessy:2011nwl}, its EM radiation in the radio band is expected to be detected by SKA, thus its host galaxy could be uniquely identified. Here, $F^{\rm SKA}_{\rm min}=\nu_{\rm SKA}F^{\rm SKA}_{\nu,{\rm min}}$ is the detector's flux limit, with $\nu_{\rm SKA}\simeq1.4~{\rm GHz}$ and $F^{\rm SKA}_{\nu,{\rm min}}\simeq1~\mu{\rm Jy}$. It should be noted that we assume that the radio radiation is isotropic, according to Ref.~\cite{Tamanini:2016zlh}. Actually, the synchrotron emission within the jet is beamed along the jet, which makes it impossible to detect the events with jets being not towards the earth, thus reducing the number of GW-EM events. However, collimation also implies a larger flux for a given source luminosity, which makes some intrinsically fainter sources being observable, thus increasing the number of GW-EM events. Because these two factors have opposite effects on the number of GW-EM events, and may counteract each other, we assume that the radiation is isotropic for the purpose of simplification.

Only the radio identification cannot complete the measurement of the redshift, so the optical/IR facilities are needed to observe the spectral features to obtain the redshift.  The host galaxy's luminosity in the $K$-band, $L_k$, is computed by converting the host total stellar mass into luminosity \cite{Tamanini:2016zlh}. According to the results of Ref.~\cite{Bruzual:2003tq}, for young stellar populations at moderate redshift, the mass-to-light ratio $M/L_k$ falls in the range $0.01-0.05$. We assume a fiducial $M/L_k = 0.03$ in the simulation.
By converting $L_k$ into apparent magnitude $m_{\rm gal}$, we assume that the redshifts of MBHB merger events that satisfy the following relationship can be measured by ELT,
\begin{align}
m_{\rm gal}=82.5-\frac{5}{2}{\rm log}_{10}\left(\frac{L_k}{3.02}\frac{{\rm s}}{{\rm erg}}\right)+5{\rm log}_{10}\left(\frac{d_{\rm L}}{{\rm pc}}\right)\leq m_{{\rm ELT}},
\end{align}
with the detection threshold $m_{{\rm ELT}}$ being set to 31.3, which is the photometric limiting magnitude of ELT corresponding to $J$-band and $H$-band \cite{Davies:2010fv}. In principle, the detection threshold should be set to 30.2 that is the limiting magnitude of $K$-band, because the host galaxy's luminosity in $K$-band is used to calculate apparent magnitude. Actually, MICADO (Multi-AO Imaging Camera for Deep Observations) on ELT will cover the wavelength range of 1000--2400 nm ($J$-band to $K$-band), so we simply choose the highest limiting magnitude, 31.3, as the detection threshold. This simplification may lead to a very small overestimation of the detection threshold, but will have no an obvious effect on the number of GW-EM events and the cosmological analysis.
Based on the above criterion, we can pick out the GW events whose redshifts can be determined. For the redshift error $\sigma_{d_{\rm L}}^{\rm reds}$, the authors of Ref.~\cite{Tamanini:2016zlh} take into account it for the GW-EM events satisfying $27.2<m_{\rm gal}<31.3$, with 27.2 being the spectroscopy limiting magnitude of ELT. In Ref.~\cite{Speri:2020hwc}, the authors use a more simple method, namely taking into account the redshift error for all the GW-EM events of $z>2$. Namely, it is assumed that the redshifts of GW events with $z<2$ are measured by spectroscopy, while those with $z>2$ are measured by photometry, because the spectroscopic redshift in the range of $z>2$ is usually unavailable \cite{Dahlen:2013fea,Speri:2020hwc}. In this work, we adopt the simplified method in Ref.~\cite{Speri:2020hwc}.
Actually, one can see from Figure~\ref{sirens} that most GW-EM events are at $z>2$. Thus, in our analysis, the redshift error is actually considered for most data points.

\subsection{{Standard siren catalog}}\label{sec:catalog}
Based on the methods discussed in the previous sections, we construct the standard siren catalogs in preparation for cosmological parameter constraints. In this work, we discuss three population models of MBHB \cite{Madau:2001sc,Volonteri:2007ax}, i.e., pop III, Q3d, and Q3nod, which are proposed \cite{Klein:2015hvg} for the simulation of standard sirens, according to the birth mechanism of MBHs and whether there exists a delay between the mergers of MBHB and their host galaxies. For the total numbers of the MBHB merger events within 5 years, we estimate them based on  the event rates given in Table I of Ref.~\cite{Klein:2015hvg}. Specifically, the numbers are 877, 41, and 610 for pop III, Q3d, and Q3nod, respectively.
For the redshift distribution and the mass distribution of MBHBs, we give numerical fitting formulas of the curves shown in Figure 3 of Ref.~\cite{Klein:2015hvg}.
The predicted MBHB merger rates as functions of redshift are given by
\begin{align}
R(z)_{{\rm pop~III}}=&\begin{cases}
2.11 z, & 0\leq z\leq 9, \\
-1.8z+35.2, & 9<z\leq 19, \\
\end{cases}\\
R(z)_{\rm{Q3d}}=&\begin{cases}
0.43z, & 0\leq z\leq 3.5, \\
-0.18z+2.12, & 3.5<z\leq 12,\\
\end{cases}\\
R(z)_{\rm{Q3nod}}=&\begin{cases}
1.67z, & 0\leq z\leq 6, \\
-0.69z+14.62, & 6<z\leq 19.
\end{cases}
\end{align}
The predicted MBHB merger rates as functions of the total redshifted mass, $M_z$, are given by
\begin{align}
R(M_z)_{{\rm pop~III}}=&\begin{cases}
10^{-10.6}\left(\frac{M_z}{M_{\odot}}\right)^{3.2}, & 10^{3}M_{\odot}\leq M_z\leq 10^{4}M_{\odot}, \\
10^{5.32}\left(\frac{M_z}{M_{\odot}}\right)^{-0.78}, & 10^{4}M_{\odot}<M_z\leq 10^{8}M_{\odot}, \\
\end{cases}\\
R(M_z)_{\rm{Q3d}}=&\begin{cases}
10^{-4.4}\left(\frac{M_z}{M_{\odot}}\right)^{{0.81}}, & 10^{4}M_{\odot}\leq M_{z}\leq 10^{6.3}M_{\odot}, \\
10^{6.65}\left(\frac{M_z}{M_{\odot}}\right)^{-0.94}, & 10^{6.3}M_{\odot}<M_{z}\leq 10^{8}M_{\odot},
\end{cases}\\
R(M_z)_{\rm{Q3nod}}=&\begin{cases}
10^{-5.44}\left(\frac{M_z}{M_{\odot}}\right)^{1.2}, & 10^4M_{\odot}\leq M_{z}\leq 10^{6.2}M_{\odot}, \\
10^{9.75}\left(\frac{M_z}{M_{\odot}}\right)^{-1.25}, & 10^{6.2}M_{\odot}<M_{z}\leq 10^{8}M_{\odot},
\end{cases}
\end{align}
with $M_z=M(1+z)$. $R(z)$ and $R(M_z)$ are in units of ${\rm yr}^{-1}$.
The sky location ($\theta$, $\phi$), the binary inclination $\iota$, the polarization angle $\psi$, and the coalescence phase $\phi_c$ are randomly chosen in the ranges of $[0,\pi]$, $[0,2\pi]$, $[0,\pi]$, $[0,2\pi]$, and $[0,2\pi]$, respectively.

\begin{figure}
  \centering
    \includegraphics[width=7.2cm,height=5.4cm]{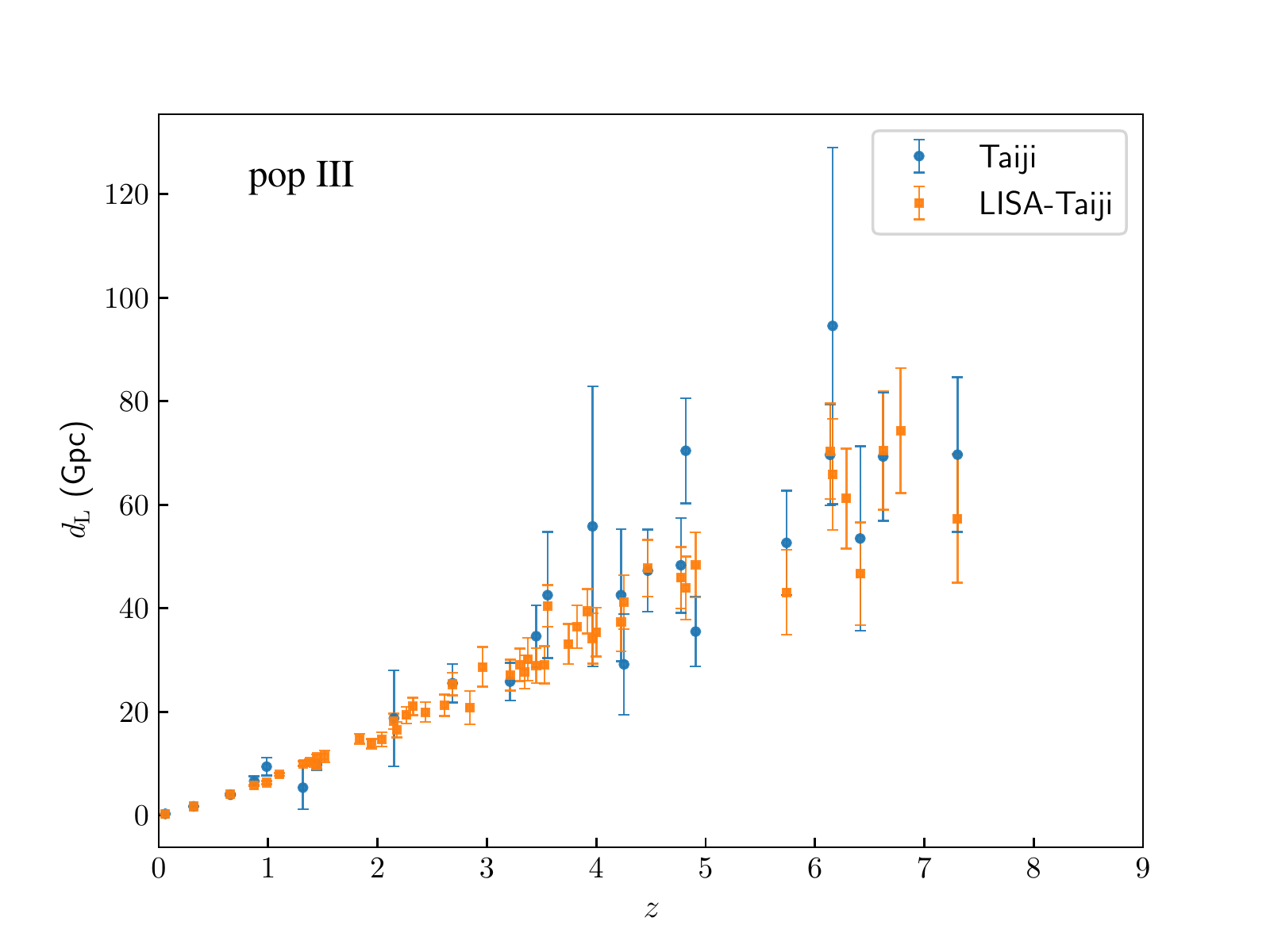}
    \includegraphics[width=7.2cm,height=5.4cm]{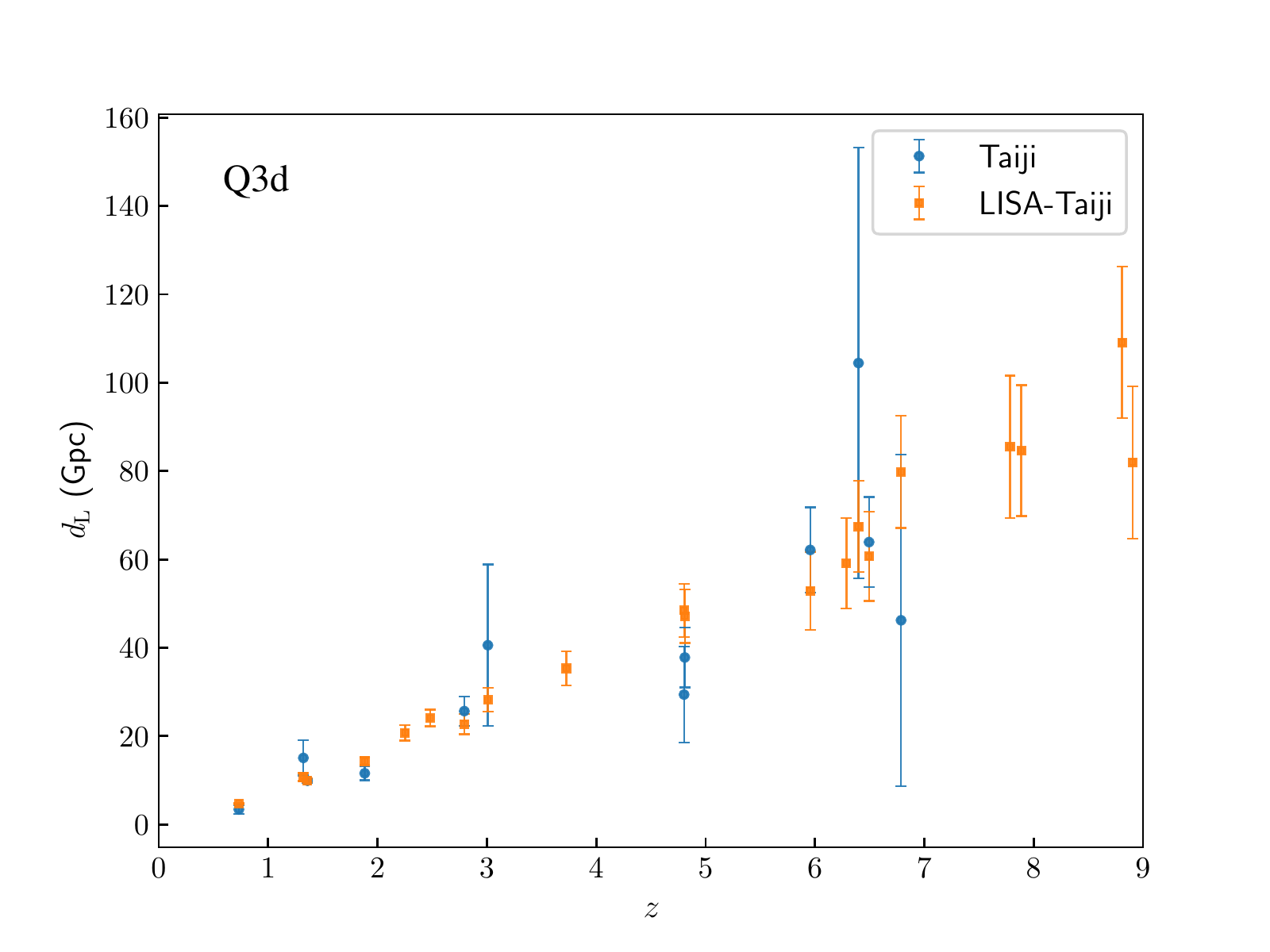}
    \includegraphics[width=7.2cm,height=5.4cm]{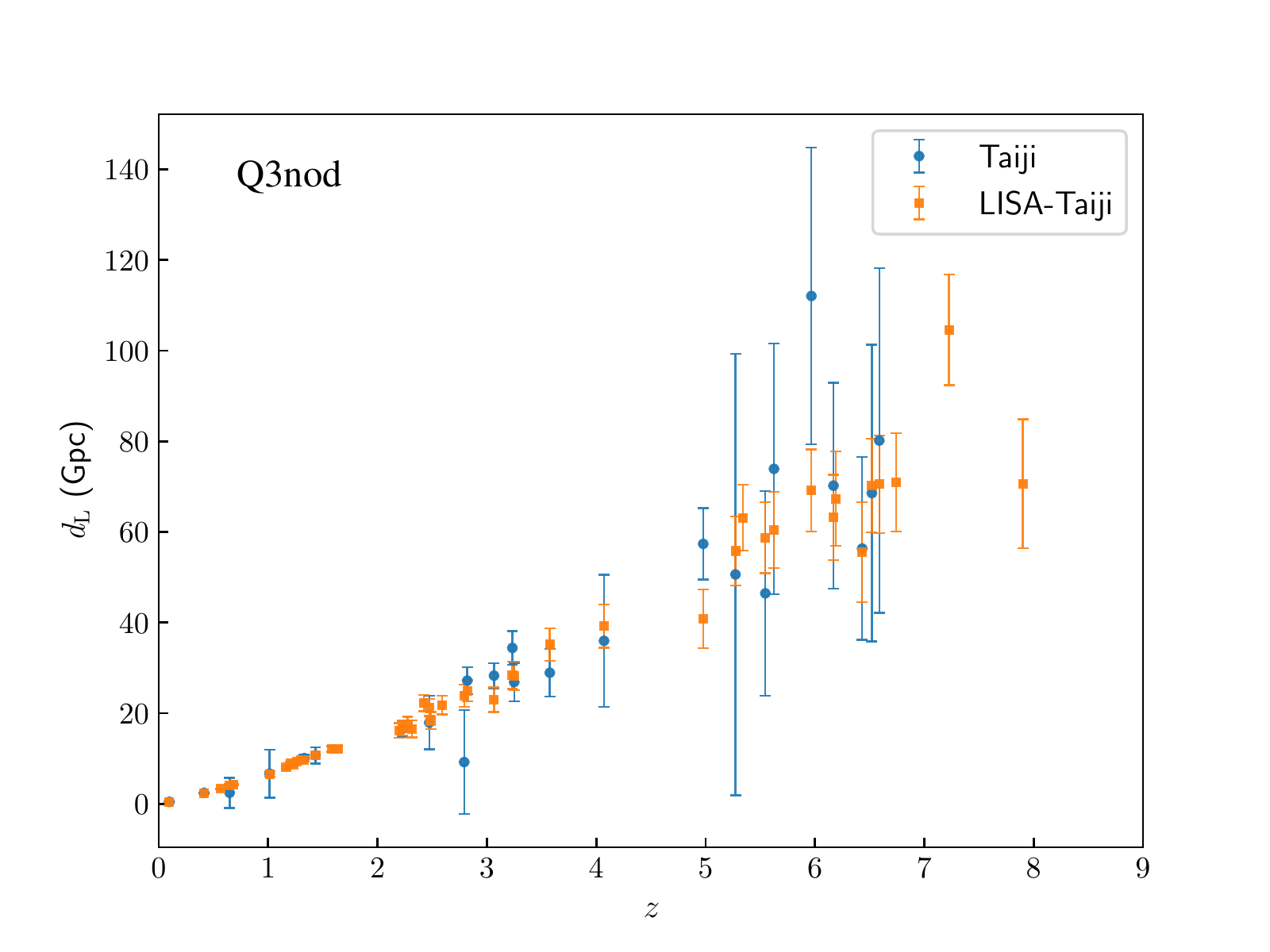}
\caption{The standard siren catalogs simulated from Taiji and the LISA-Taiji network within 5-year operation time based on the pop III, Q3d, and Q3nod models.} \label{sirens}
\end{figure}
After selecting the GW events satisfying SNR $>$ 8 and $\Delta\Omega<10~{\rm deg}^2$, and further selecting the GW events whose redshifts can be detected by SKA and ELT, we obtain the useful standard sirens. For each standard siren, we calculate its luminosity distance $d_{\rm L}$ and the error of luminosity distance $\sigma_{d_{\rm L}}$. The fiducial values of cosmological parameters are set to the best-fit values of the {\it Planck} 2018 results \cite{Aghanim:2018eyx}. Then, for each MBHB model, we can construct a standard siren catalog including the redshift $z$, luminosity distance $d_{\rm L}$, and error of luminosity distance $\sigma_{d_{\rm L}}$ of MBHBs.

We show the simulated standard sirens in Figure \ref{sirens}. Firstly, it is found that the numbers of standard sirens detected by the LISA-Taiji network are much more than those detected by the single Taiji mission, which also can be seen in Table \ref{tab1}. For each MBHB model, the LISA-Taiji network approximately doubles the detection number compared to the single Taiji mission. This is due to the fact that the network could improve SNR and the location accuracy of the GW event.
Secondly, the network can detect the GW events at higher redshifts due to the improvement of SNR. For example, from the middle panel of Figure \ref{sirens}, we can see that for the Q3d model, the redshifts of the standard sirens detected by the network could reach $z\sim9$. Thirdly, the LISA-Taiji network can improve the measurement accuracy of luminosity distance to some extent, which can be seen from the error bars in Figure \ref{sirens}. These improvements make us expect that the LISA-Taiji network can greatly improve the capability of constraining cosmological parameters.

\begin{table}[htbp]
\footnotesize
\centering
\setlength\tabcolsep{10pt}
\renewcommand{\arraystretch}{1.5}
\caption{\label{tab1} {The numbers of the standard sirens simulated from Taiji and the LISA-Taiji network within 5-year operation time, based on
the pop III, Q3d, and Q3nod models of MBHB population.}}
\begin{tabular}{c c c c}
\hline
\multirow{2}{*}{Model}&\multicolumn{3}{c}{Number of standard sirens}\\
\cline{2-4}
&&Taiji&network\\
\hline
pop III&&25&50\\
Q3d&&12&20\\
Q3nod&&24&44\\
\hline
\end{tabular}
\centering
\end{table}

\section{Cosmological parameter estimation}\label{sec:Result}
In this section, we shall report the constraint results of cosmological parameters. In theory, the luminosity distance $d_{\rm L}$ of a GW source at redshift $z$ is determined by a specific cosmological model. The $\Lambda$CDM model [$w(z) = -1$], the $w$CDM model [$w(z) = \rm{constant}$], and the Chevallier--Polarski--Linder (CPL) model [$w(z)=w_{\rm{0}}+w_{\rm{a}}z/(1+z)$] \cite{Chevallier:2000qy,Linder:2002et} are considered in this paper. For the CMB data, we employ the ``{\it Planck} distance priors'' from the {\it Planck} 2018 observation \cite{Chen:2018dbv}. We use $\sigma(\xi)$ and $\varepsilon(\xi)$ to represent the absolute error and the relative error of the parameter $\xi$, respectively, with $\varepsilon(\xi)$ defined as $\varepsilon(\xi)=\sigma(\xi)/\xi$.

\begin{table*}[htbp]
\small
\setlength\tabcolsep{6pt}
\renewcommand{\arraystretch}{1.5}
\caption{\label{tab2} {The absolute errors ($1\sigma$) and the relative errors of the cosmological parameters in the $\Lambda$CDM, $w$CDM, and CPL models using the mock data of the LISA-Taiji network. Here, $\sigma(\xi)$ and $\varepsilon(\xi)$ represent the absolute and relative errors of the parameter $\xi$, respectively. Note also that $\varepsilon(\xi)$ is defined as $\varepsilon(\xi)=\sigma(\xi)/\xi$, and $H_0$ is in units of ${\rm km\ s^{-1}\ Mpc^{-1}}$.}}
\centering
\begin{tabular}{ccccccccccccccc}
\hline\multirow{2}{*}{Error} &\multicolumn{3}{c}{$\Lambda$CDM}&& \multicolumn{3}{c}{$w$CDM}&& \multicolumn{3}{c}{CPL}\\
 \cline{2-4}\cline{6-8}\cline{10-12}
 & pop III &Q3d  &Q3nod& &pop III&Q3d &Q3nod &&pop III&Q3d& Q3nod\\
\hline
$\sigma(\Omega_{\rm m})$
                   & $0.025$
                   & $0.073$
		          &$0.027$&
                   & $0.036$
                   & $0.078$
		          &$0.037$&
                   & $0.054$
                   & $0.099$
		          &$0.056$\\

$\sigma(H_0)$
                   & $0.86$
                 & $3.25$
		           &$0.94$&
                   & $1.85$
                   & $10.10$
		          &$1.75$&
                   & $2.40$
                   & $10.00$
		          &$2.10$\\

$\sigma(w)$
                    & $-$
                    & $-$
		             &$-$&
                   & $0.245$
                    & $0.735$
		          &$0.230$&
                    & $-$
                   & $-$
		            &$-$&\\

$\sigma(w_0)$
                    & $-$
                   & $-$
		          &$-$&
                   & $-$
                   & $-$
		           &$-$&
                   & $0.380$
                 & $0.965$
		          &$0.340$&\\

$\sigma(w_a)$
                   & $-$
                   & $-$
		          &$-$&
                   & $-$
                   & $-$
		          &$-$&
                 & $1.95$
                   & $-$
		          &$2.05$&\\\hline
$\varepsilon(\Omega_{\rm m})$
                   & $0.078$
                   & $0.212$
		           &$0.084$&
                   & $0.113$
                   & $0.258$
		           &$0.114$&
                   & $0.165$
                   & $0.304$
		           &$0.171$\\

$\varepsilon(H_0)$
                   & $0.013$
                    & $0.049$
		          &$0.014$&
                   & $0.027$
                   & $0.141$
		          &$0.026$&
                   & $0.036$
                   & $0.141$
		          &$0.031$\\

$\varepsilon(w)$
                    & $-$
                   & $-$
		            &$-$&
                   & $0.229$
                    & $0.525$
		          &$0.215$&
                    & $-$
                   & $-$
		          &$-$&\\

$\varepsilon(w_0)$
                  & $-$
                   & $-$
		           &$-$&
                   & $-$
                   & $-$
		          &$-$&
                   & $0.409$
                 & $0.832$
		          &$0.378$&\\

\hline
\end{tabular}
\centering
\end{table*}

\begin{table*}[htbp]
\setlength\tabcolsep{1.0pt}
\renewcommand{\arraystretch}{1.5}
\caption{\label{tab3} {The absolute errors ($1\sigma$) and the relative errors of the cosmological parameters in the $\Lambda$CDM, $w$CDM, and CPL models using the CMB, Taiji(Q3nod), network(Q3nod), CMB+Taiji(Q3nod), and CMB+network(Q3nod) data. Here, $\sigma(\xi)$ and $\varepsilon(\xi)$ represent the absolute and relative errors of the parameter $\xi$, respectively. Note also that $\varepsilon(\xi)$ is defined as $\varepsilon(\xi)=\sigma(\xi)/\xi$, and $H_0$ is in units of ${\rm km\ s^{-1}\ Mpc^{-1}}$.}}
\centering
\resizebox{\textwidth}{!}
{
\begin{tabular}{ccccccccccccccccccc}
\hline\multirow{2}{*}{Error} &\multicolumn{5}{c}{$\Lambda$CDM}&& \multicolumn{5}{c}{$w$CDM}&& \multicolumn{5}{c}{CPL}\\
 \cline{2-6}\cline{8-12}\cline{14-18}
  &CMB & Taiji &network& CMB+Taiji &CMB+network&&CMB & Taiji &network &CMB+Taiji &CMB+network&&CMB & Taiji &network &CMB+Taiji &CMB+network\\
\hline
$\sigma(\Omega_{\rm m})$
                   & $0.009$
                   & $0.068$
		           &$0.027$
                   & $0.008$
                   & $0.007$&
                    & $0.057$
		           &$0.087$
                   & $0.037$
		           &$0.026$
                    &$0.010$&
                    &$0.059$
                    &$0.086$
                    &$0.056$
                    &$0.032$
                    &$0.018$\\

$\sigma(H_0)$
                   & $0.61$
                   & $2.60$
		           &$0.94$
                   & $0.58$
                   & $0.46$&
                   & $6.15$
		           &$3.90$
                   & $1.75$
		           &$2.80$
                    &$1.00$&
                    &$6.25$
                    &$4.05$
                    &$2.10$
                    &$3.25$
                    &$1.90$\\

$\sigma(w)$
                   & $-$
                   & $-$
		           &$-$
                   & $-$
                   & $-$&
                   & $0.215$
		           &$0.695$
                   & $0.230$
		           &$0.097$
                    &$0.042$&
                    &$-$
                    &$-$
                    &$-$
                    &$-$
                    &$-$\\

$\sigma(w_0)$
                   & $-$
                   & $-$
		           &$-$
                   & $-$
                   & $-$&
		           &$-$
                   & $-$
                   & $-$
		           &$-$
                    &$-$&
                    &$0.575$
                    &$0.885$
                    &$0.340$
                    &$0.510$
                    &$0.230$\\

$\sigma(w_a)$
                   & $-$
                   & $-$
		           &$-$
                   & $-$
                   & $-$&
		           &$-$
                   & $-$
                   & $-$
		           &$-$
                    &$-$&
                    &$-$
                    &$-$
                    &$2.05$
                    &$1.80$
                    &$0.67$\\
\hline

$\varepsilon(\Omega_{\rm m})$
                   & $0.027$
                   & $0.201$
		           &$0.084$
                   & $0.026$
                   & $0.020$&
                   & $0.178$
		           &$0.263$
                   & $0.114$
		           &$0.082$
                    &$0.030$&
                    &$0.183$
                    &$0.246$
                    &$0.171$
                    &$0.097$
                    &$0.056$\\

$\varepsilon(H_0)$
                    & $0.009$
                   & $0.039$
		           &$0.014$
                   & $0.009$
                   & $0.007$&
                   & $0.090$
		           &$0.057$
                   & $0.026$
		           &$0.041$
                    &$0.015$&
                    &$0.092$
                    &$0.059$
                    &$0.031$
                    &$0.049$
                    &$0.028$\\

$\varepsilon(w)$
                   & $-$
                   & $-$
		           &$-$
                   & $-$
                   & $-$&
                   & $0.211$
		           &$0.489$
                   & $0.215$
		           &$0.097$
                    &$0.042$&
                    &$-$
                    &$-$
                    &$-$
                    &$-$
                    &$-$\\

$\varepsilon(w_0)$
                   & $-$
                   & $-$
		           &$-$
                   & $-$
                   & $-$&
		           &$-$
                   & $-$
                   & $-$
		           &$-$
                    &$-$&
                    &$0.871$
                    &$0.651$
                    &$0.378$
                    &$0.823$
                    &$0.242$\\

\hline
\end{tabular}
}
\centering
\end{table*}

\begin{figure}[htbp]
\begin{center}
\includegraphics[width=7.2cm,height=5.4cm]{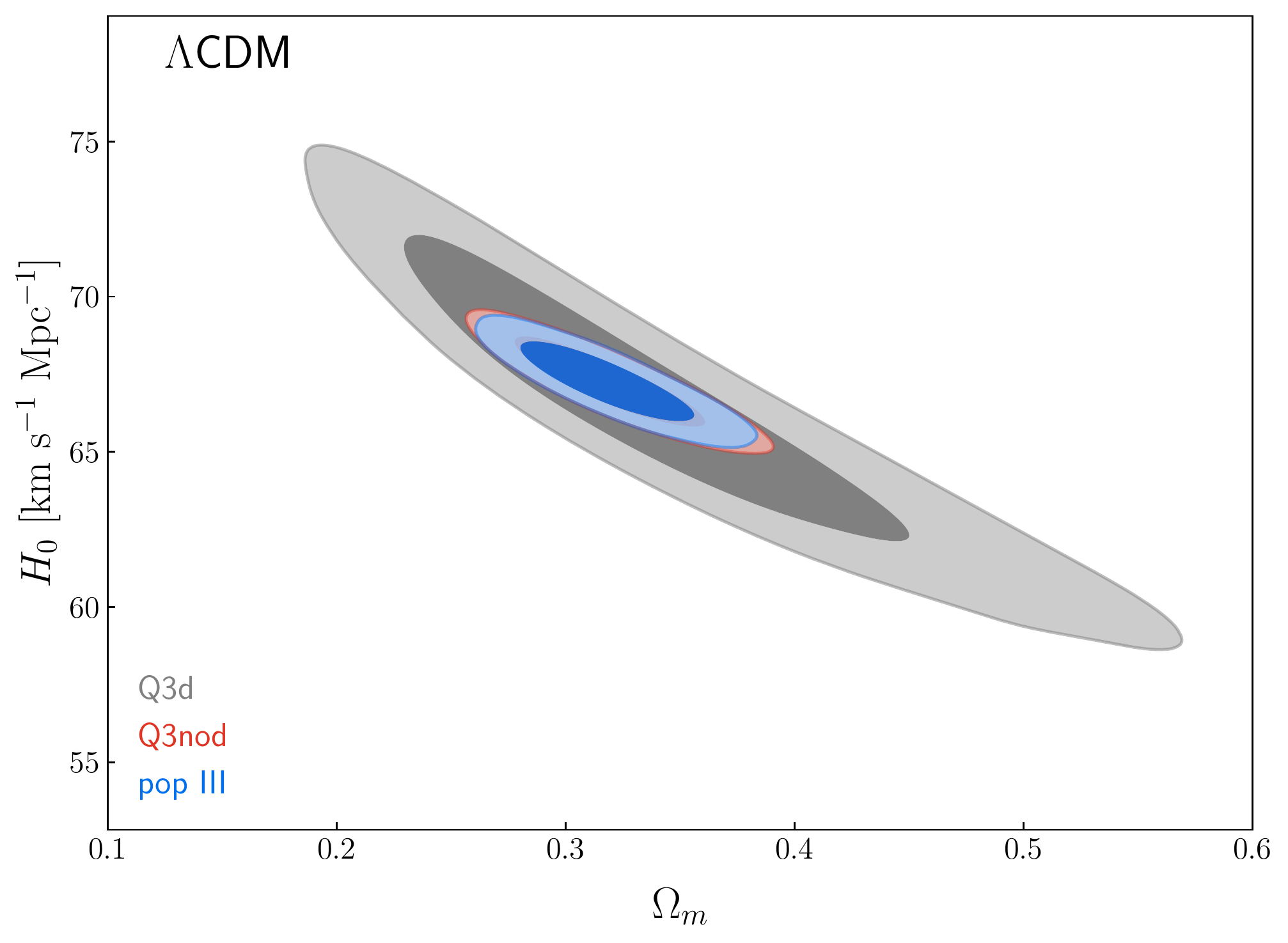} \\
\end{center}
\caption{{The two-dimensional marginalized contours (68.3\% and 95.4\% confidence level) in the $\Omega_{m}$--$H_{0}$ plane considering three MBHB models for the $\Lambda$CDM model.}} \label{models}
\end{figure}

Firstly, let us take the simplest $\Lambda$CDM model as an example to discuss the constraint results for the three MBHB models. From Figure \ref{models}, we find that the pop III model gives the best constraints because it has the maximal event number among the three MBHB models. The Q3nod model is similar with the pop III model, but the Q3d model gives the worst constraints because it has the fewest event number. Quantitatively, the pop III model gives the relative errors $\varepsilon(\Omega_{\rm m})=7.8\%$ and $\varepsilon(H_0)=1.3\%$, the Q3nod model gives $\varepsilon(\Omega_{\rm m})=8.4\%$ and $\varepsilon(H_0)=1.4\%$, and the Q3d model gives $\varepsilon(\Omega_{\rm m})=21.2\%$ and $\varepsilon(H_0)=4.9\%$, as shown in Table \ref{tab2}. Here, we notice that solely using the standard sirens from the LISA-Taiji network could provide a tight constraint on $H_0$, with the precision close to 1\%, which is the standard of precision cosmology.
Compared with the single Taiji mission, the LISA-Taiji network reduces the absolute errors of $\Omega_{\rm m}$ and $H_0$ by 60.3\% and 63.8\%, respectively, which can be seen from Table \ref{tab3}. This indicates that the GW detection network could provide a more accurate measurement of $H_0$ than a single detector, which is helpful for solving the Hubble constant tension.

\begin{figure}[htbp]
\begin{center}
\includegraphics[width=7.2cm,height=5.4cm]{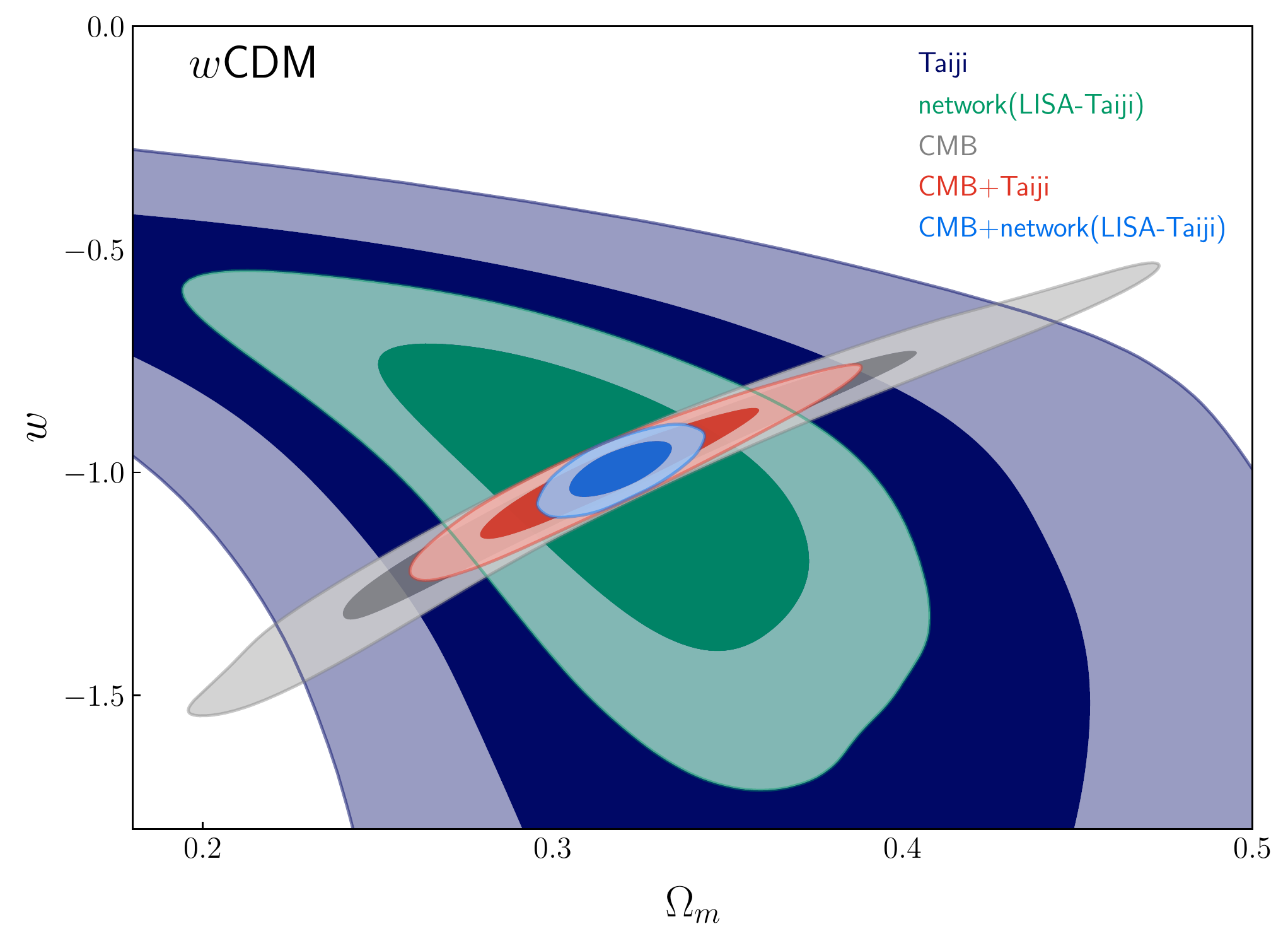} \\
\includegraphics[width=7.2cm,height=5.4cm]{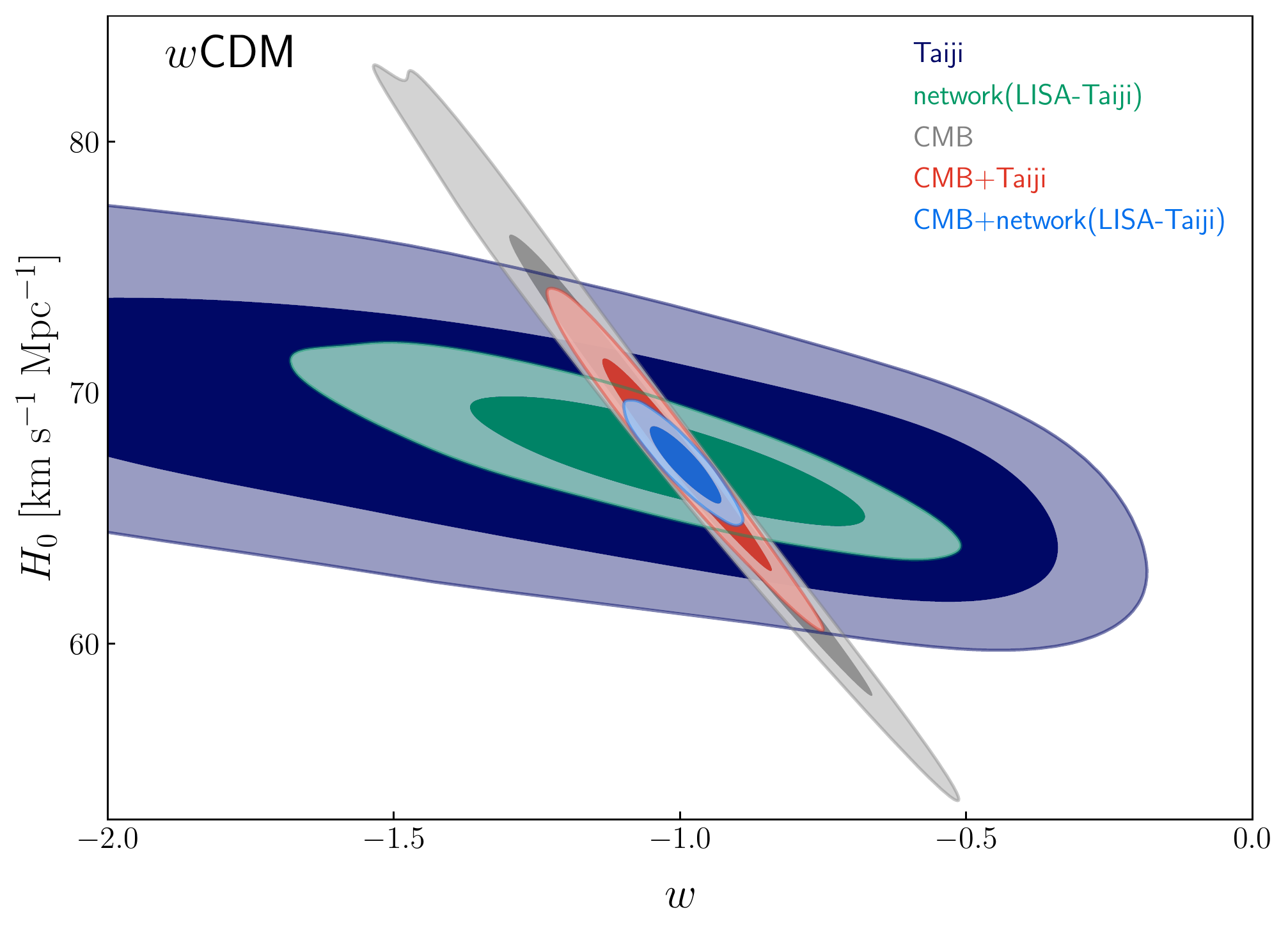}
\end{center}
\caption{{The two-dimensional marginalized contours (68.3\% and 95.4\% confidence level) in the $\Omega_{\rm m}-w$ and $w-H_0$ planes for the $w$CDM. The mock data of Taiji and the LISA-Taiji network are simulated based on the Q3nod model.}} \label{cosmodels}
\end{figure}

Now let us have a look at the improvements of the constraints on the EoS parameter of dark energy. We first discuss the $w$CDM model that has only one dark-energy EoS parameter. In Figure~\ref{cosmodels}, we show the two-dimensional posterior contours in the $\Omega_{\rm m}-w$ and $w-H_0$ planes using the data of Taiji, network, CMB, CMB+Taiji, and CMB+network. From the purple contour and the green contour, we can see that the network can give much better constraints than the single Taiji mission. The gray contour represents the constraint from the CMB data, and this contour is almost orthogonal to the green contour, which indicates that the standard siren data could significantly break the parameter degeneracies. By comparing the red contour and the blue contour, we can see that, although both the single Taiji mission and the network can break degeneracies, the network can improve the constraint accuracies much better. As we mentioned above, this is because the LISA-Taiji network can detect more GW-EM events than Taiji, and can detect the events at higher redshifts. In addition, the network also improves SNR, thus reducing the error of $d_{\rm L}$.
Concretely, the LISA-Taiji network could reduce the absolute error of $w$ by 66.9\% compared with the single Taiji mission. The data combination CMB+network could reduce the absolute error of $w$ by 56.7\% compared with CMB+Taiji. It is worth emphasizing that using the CMB+network data, the constraint precision of $w$ could reach $4.2\%$, which is comparable with the result of {\it Planck} 2018 TT,TE,EE+lowE+lensing+SNe+BAO \cite{Aghanim:2018eyx}. Note also that here we use only the {\it Planck} distance priors, but not the {\it Planck} full data of CMB power spectra.

For the CPL model that has two dark-energy EoS parameters $w_0$ and $w_a$, we also find that the LISA-Taiji network data give better constraints compared with the single Taiji mission. The detailed results are shown in Table \ref{tab3}. For the parameter $w_0$, Taiji gives the relative error $\varepsilon(w_0)=65.1\%$, and the LISA-Taiji network gives the relative error $\varepsilon(w_0)=37.8\%$. Compared with the single Taiji mission, the LISA-Taiji network can reduce the absolute error of $w_0$ by 61.6\%. For the parameter $w_a$, Taiji cannot constrain $w_a$ well, but the LISA-Taiji network gives the absolute error $\sigma(w_a)=2.05$. When the CMB data are combined, CMB+Taiji gives $\varepsilon(w_0)=82.3\%$ and $\sigma(w_a)=1.80$, and CMB+network gives $\varepsilon(w_0)=24.2\%$ and $\sigma(w_a)=0.67$. Compared with CMB+Taiji, CMB+network could reduce the absolute errors of $w_0$ and $w_a$ by 41.5\% and 54.9\%, respectively. This indicates that the future LISA-Taiji network combined with the CMB observation will play an important role in constraining the EoS parameter of dark energy.

It should be noted that we make some assumptions and approximations in this work. For the redshift distribution and the mass distribution of MBHBs, we use the numerical fitting formulas. For the redshift measurement,
we assume that the redshifts of GW events satisfying $z>2$ are measured by photometry, so we take into account the redshift errors for these events. In addition, we simply assume that the radio radiation in the EM counterpart is isotropic. We have discussed the rationalities of these assumptions in the previous sections. The main purpose of this work is to make a preliminary forecast on the capability of the LISA-Taiji network of improving the estimations of cosmological parameters, compared with a single GW detector. For this purpose, these assumptions do not affect our main conclusions. For a more specific investigation on these issues, we leave it for future works.

\section{Conclusion}\label{sec:con}

In this work, we forecast the capability of the future space-based GW detection network to constrain cosmological parameters. We consider a detection network composed of the European LISA mission and the Chinese Taiji mission. The configuration angle between the LISA and Taiji is considered to be $40^{\circ}$. Three models for MBHB, i.e., pop III, Q3d, and Q3nod, are used to simulate the EM counterpart detection and estimate the number of GW-EM events.
Three typical cosmological models, i.e., the $\Lambda$CDM, $w$CDM, and CPL models, are chosen as representatives.

We find that the LISA-Taiji network could significantly improve the constraint accuracies of cosmological parameters compared with the single Taiji mission. This is mainly due to three aspects: (i) the LISA-Taiji network could increase the number of the GW-EM events; (ii) the LISA-Taiji network could detect the GW events at higher redshift; and (iii) the LISA-Taiji network could reduce the error of luminosity distance. Taking the Q3nod model as an example, the LISA-Taiji network increases the GW-EM number from 24 to 44, compared with the singe Taiji mission. The redshifts of GW events detected by the LISA-Taiji network can reach to $z\sim8$. The errors of luminosity distances are also reduced to some extent, which can be seen from the error bars in Figure \ref{sirens}.

For the simplest $\Lambda$CDM model, we find that the pop III model gives the best constraints due to its maximal event number among these three MBHB models. The Q3nod model is similar with pop III, but the Q3d model gives the worst constraints due to its fewest event number. The pop III model gives the relative errors $\varepsilon(\Omega_{\rm m})=7.8\%$ and $\varepsilon(H_0)=1.3\%$; the Q3nod model gives $\varepsilon(\Omega_{\rm m})=8.4\%$ and $\varepsilon(H_0)=1.4\%$; the Q3d model gives $\varepsilon(\Omega_{\rm m})=21.2\%$ and $\varepsilon(H_0)=4.9\%$. So, we find that solely using the standard sirens from the LISA-Taiji network, the constraint precision of $H_0$ could reach about 1\%, which is the standard of precision cosmology. In addition, compared with the single Taiji mission, the LISA-Taiji network could reduce the relative error of $H_0$ by 63.8\%. This indicates that the LISA-Taiji network is helpful in addressing the $H_0$ tension.


For the dark-energy EoS parameter, we first discuss it in the $w$CDM model. Taking the Q3nod model as an example, we find that the LISA-Taiji network reduces the absolute error of $w$ by 66.9\% compared with the single Taiji mission. We also see that the LISA-Taiji network improves the capability of breaking the parameter degeneracies inherent in the CMB data. Specifically, the CMB+network data reduces the absolute error of $w$ by 56.7\% compared with CMB+Taiji.
The constraint precision of $w$ could reach $4.2\%$ using the CMB+network data, which is comparable with the result of $Planck$ 2018 TT,TE,EE+lowE+lensing+SNe+BAO.
For the CPL model that has two dark-energy EoS parameters $w_0$ and $w_a$, the LISA-Taiji network can reduce the absolute error of $w_0$ by 61.6\%, compared with the single Taiji mission. For the parameter $w_a$, Taiji cannot constrain $w_a$ well, but the LISA-Taiji network can give the absolute error $\sigma(w_a)=2.05$. Compared with CMB+Taiji, CMB+network could reduce the absolute errors of $w_0$ and $w_a$ by 41.5\% and 54.9\%, respectively. This indicates that the future LISA-Taiji network combined with the CMB observation will play an important role in constraining the EoS parameters of dark energy.

In the next few decades, GW detectors are expected to form powerful detection networks. This allows researchers to use multiple detectors to jointly detect a GW source, thereby improving the measurement accuracies of source parameters. At the same time, the detections of different GW sources in multiple GW frequency bands can help us test the theory of gravity, and can also lead to a more comprehensive understanding of the properties of compact objects and the evolution history of the universe. We expect that the LISA-Taiji network could play an important role in these aspects, and we will further discuss them in depth in our future work.

\begin{acknowledgments}
{We are very grateful to Antoine Klein, Alberto Mangiagli, Alberto Sesana, and Nicola Tamanini for fruitful discussions on the identifications of EM counterparts, and also grateful to Tao Yang, Wen-Hong Ruan, and Ze-Wei Zhao for discussions on the detections of GW sources.}
This work was supported by the National Natural Science Foundation of China (Grants Nos. 11975072, 11835009, 11875102, and 11690021), the Liaoning Revitalization Talents Program (Grant No. XLYC1905011), the Fundamental Research Funds for the Central Universities (Grant No. N2005030), and the National Program for Support of Top-Notch Young Professionals (Grant No. W02070050).
\end{acknowledgments}

\bibliography{network}

\end{document}